\documentclass[twocolumn,secnumarabic,amssymb,superscriptaddress,nobibnotes, aps, prd]{revtex4-2}

\usepackage{comment}
\usepackage{graphicx}
\usepackage[charter,greekuppercase=italicized]{mathdesign}
\usepackage{amsmath, amsfonts, amssymb}
\usepackage{hyperref}
\usepackage{xcolor}
\hypersetup{
    colorlinks = true,
    linkcolor = blue,
    anchorcolor = blue,
    citecolor = blue,
    filecolor = blue,
    urlcolor = blue
    }   
\usepackage{multirow}
\usepackage{hhline}

\begin{document}
\title{Evidence of quantum spin liquid state in a Cu$^{2+}$-based $S = 1/2$ triangular lattice antiferromagnet}%

\author{K. Bhattacharya}
\address{Department of Physics, Shiv Nadar Institution of Eminence, Gautam Buddha Nagar, UP 201314, India}

\author{S. Mohanty}
\address{School of Physics, Indian Institute of Science Education and Research, Thiruvananthapuram 695551, India}

\author{A. D. Hillier}
\address{ISIS Facility, STFC Rutherford Appleton Laboratory, Chilton, Oxfordshire OX11 0QX, United Kingdom}
\author{M. T. F. Telling}
\address{ISIS Facility, STFC Rutherford Appleton Laboratory, Chilton, Oxfordshire OX11 0QX, United Kingdom}

\author{R. Nath}
\address{School of Physics, Indian Institute of Science Education and Research, Thiruvananthapuram 695551, India}

\author{M. Majumder}
\email[Corresponding author: ]{mayukh.majumder@snu.edu.in}
\address{Department of Physics, Shiv Nadar Institution of Eminence, Gautam Buddha Nagar, UP 201314, India}

\date{\today}
\begin{abstract}
The layered triangular lattice owing to $1:2$ order of $B$ and $B'$ sites in the triple perovskite $A_3 B B'_2$O$_9$ family provides an enticing domain for exploring the complex phenomena of quantum spin liquids (QSLs). We report a comprehensive investigation of the ground state properties of Sr$_3$CuTa$_2$O$_9$ that belongs to the above family, by employing magnetization, specific heat, and muon spin relaxation ($\mu$SR) experiments down to the lowest temperature of 0.1~K. Analysis of the magnetic susceptibility indicates that the spin-lattice is a nearly isotropic $S = 1/2$ triangular lattice. We illustrate the observation of a gapless QSL, in which conventional spin ordering or freezing effects are absent, even at temperatures more than two orders of magnitude smaller than the exchange energy ($J_{\rm CW}/k_{\rm B} \simeq -5.04$~K). Magnetic specific heat in zero-field follows a power law, $C_{\rm m} \sim T^\eta$, below 1.2~K with $\eta \approx 2/3$, which is consistent with a theoretical proposal of the presence of spinon Fermi surface. Below 1.2~K, the $\mu$SR relaxation rate shows no temperature dependence, suggesting persistent spin dynamics as expected for a QSL state. Delving deeper, we also analyze longitudinal field $\mu$SR spectra revealing strong dynamical correlations in the spin-disordered ground state. All of these highlight the characteristics of spin entanglement in the QSL state.
\end{abstract}

\maketitle
\textit{Introduction:} Quantum spin liquid (QSL) is an exotic and highly entangled quantum state with no spontaneous symmetry breaking down to absolute zero temperature despite strong correlations among spins. Such a quantum phase is characterized by an emergent gauge field and fractional excitations, called spinons~\cite{Balents199,Lancaster127}. This state was first theoretically predicted by P. W. Anderson as a resonating valance bond (RVB) state for interacting Heisenberg spins in a two-dimensional (2D) triangular lattice antiferromagnet (TLAF)~\cite{Anderson153}. Subsequently, it was recognized that the true ground state of the isotropic Heisenberg TLAF is a three sub-lattice 120$^{\circ}$ N$\acute{e}$el order~\cite{PhysRevLett.60.2531,PhysRevB.40.2727}. Thereafter, considerable effort has been dedicated to stabilizing the QSL state in a Heisenberg TLAF. For instance, (i) in an isotropic TLAF with Heisenberg interactions, the competing nearest-neighbor (NN) ($J_1$) and next-nearest-neighbor (NNN) ($J_2$) interactions can stabilize a gapless Dirac QSL state for $0.08 \lesssim J_2/J_1 \lesssim 0.16$ that is sandwiched between the 120$^{\circ}$ N$\acute{e}$el order and stripe state in the phase diagram~\cite{Sherman165146,DrescherL220401}. A tiny $J_3$ can also destabilize the magnetic long-range-order (LRO) leading to a QSL phase in this model~\cite{Merino245112}. However, there are theoretical results that illustrate the importance of third nearest-neighbor exchange coupling ($J_3$) to achieve QSL state.~\cite{JiangL140411}, (ii) Considering spatially anisotropic Heisenberg interactions ($J$ and $J^\prime$), two different types of QSL states are favored: gapless QSL for $J^\prime/J\lesssim 0.65$ and gapped QSL for $0.65 \lesssim J^\prime/J \lesssim 0.8$~\cite{Yunoki014408}.

Over the years, the pursuit of new TLAFs with low spin values (e.g. $S = 1/2$) has become a focal area of research, as it is believed that QSL is a manifestation of enhanced quantum fluctuations and magnetic frustration. However, the experimental realization of QSL in $S = 1/2$ Heisenberg TLAFs is scarce due to the availability of limited model materials. Typically, the majority of the Heisenberg TLAFs show magnetic LRO at finite temperatures due to the inter-layer couplings and exchange anisotropy which are inherently present in real materials~\cite{Lal014429, Li4216,Kamiya2666, Yokota014403, Kojima174406, Ranjith094426,Ranjith014415}. To the best of our knowledge as far as inorganic compounds are concerned, QSL in a Cu$^{2+}$-based $S=1/2$ Heisenberg TLAF is realized only in   
Sr$_3$CuSb$_2$O$_9$~\cite{Kundu267202} reported to date (Note that, Ba$_3$CuSb$_2$O$_9$, likewise exhibits a QSL ground state; however, there is still disagreement over the arrangement of Cu$^{2+}$ ions, specifically whether it is triangular or Honeycomb~\cite{doi:10.1073/pnas.1508941112, PhysRevLett.106.147204, doi:10.1126/science.1212154}.). Sr$_3$CuSb$_2$O$_9$ belongs to the triple perovskite family. The family of triple perovskite ($A_3B B'_2$O$_9$ with $A$ = Sr/Ba, $B$ = Cu/Ca/Te/Os, $B'$= Sb/Ru/Ir/Fe)~\cite{Zhou147204, Wallace20344, Rijssenbeek2090, TANG86,Thakur4741} compounds are interesting  since some of the family members have $1:2$ ordering of $B/B'$ sites, owing to the site sharing of $B$ and $B'$ elements~\cite{Wallace20344,Rijssenbeek2090}. Depending on their crystal structure and space group, the $B$ site could form a superlattice structure with a specific propagation vector. For Sr$_3$CuSb$_2$O$_9$, Mahajan~\textit{et.~al.} have shown that it forms a superlattice structure with a propagation vector ($\frac{1}{3}$, $\frac{1}{3}$, $\frac{1}{3}$) and effectively forms an edge-shared triangular lattice on the (111) plane. In this pseudo-cubic system, the Cu$^{2+}$ and Sb$^{5+}$ planes are present in $1:2$ order and the compound evinces QSL state~\cite{Kundu267202}. Thus, this family of compounds holds great potential to showcase quantum magnetism due to frustration.
With this motivation, we investigated the ground state of another member of this family Sr$_3$CuTa$_2$O$_9$ (SCTO), where the $S = 1/2$ moments are embedded in an edge-shared triangular lattice. From a detailed experimental investigation by employing magnetization, specific heat, and muon spin relaxation/rotation ($\mu$SR) techniques, we provide solid evidence of a gapless QSL state in SCTO. This compound serves as a model isotropic edge-shared TLAF that hosts QSL.
\begin{figure}
\includegraphics[scale=0.35]{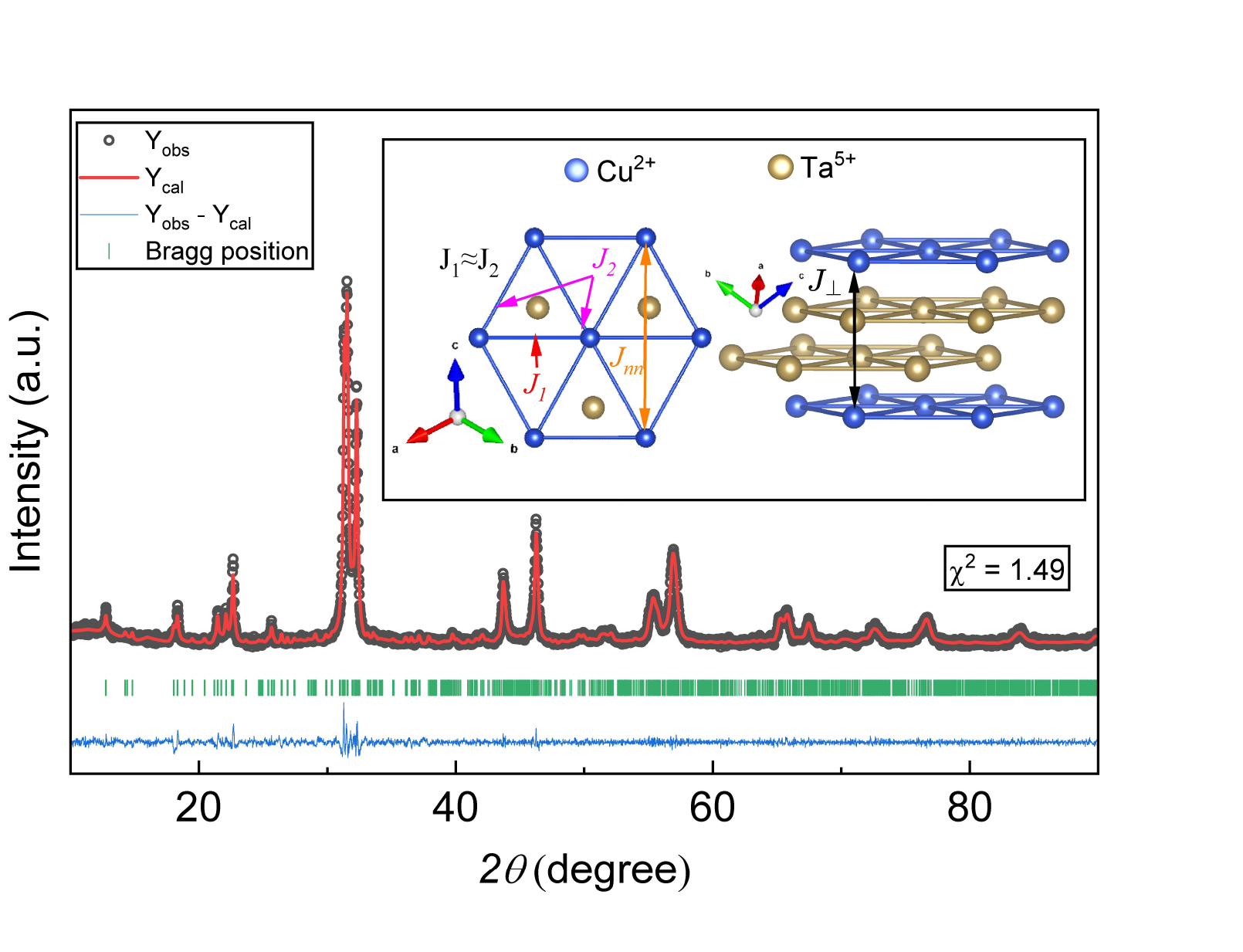}
\caption{\label{FIG1} Rietveld refinement of the powder XRD pattern taking the space group $P\bar{1}$. Inset: depicts Cu-layers arranged in a triangular lattice. The possible nearest-neighbor ($J_1$ and $J_2$), next-nearest-neighbor ($J_{nn}$), and inter-plane ($J_{\perp}$) exchange couplings are illustrated.}  
\end{figure}

\textit{Powder x-ray diffraction:} Polycrystalline SCTO sample was synthesized following the procedure described in the supplementary material (SM)~\cite{Supp}. Rietveld refinement of the powder x-ray diffraction (XRD) pattern of SCTO at room temperature was carried out assuming a tetragonal space group $P4/mmm$ (No.~123) with lattice parameters $a=b=3.93$ and $c=4.14$~\AA, which is a pseudocubical structure (details are in SM~\cite{Supp}). The best refinement was achieved with a goodness-of-fit $\chi^2 \sim 5.06$.
 Some peaks at lower angles ($2 \theta < 30^{\circ}$) couldn't be indexed with this space group as shown in SM~\cite{Supp}. These peaks correspond to a $k$-value of $\frac{1}{3}(111)$. Thus, these peaks are associated with the superlattice structure owing to the inter-site mixing of Cu$^{2+}$/Ta$^{5+}$ with occupancies $\frac{1}{3}/\frac{2}{3}$ and form a $1:2$ order, which is common in this triple perovskite systems. Note that the presence of superlattice structure is dependent on the synthesis conditions~\cite{Levin43}. Superlattice peaks along with Bragg peaks of the $P4/mmm$ space group are well fitted with a lower symmetry triclinic space group $P\bar{1}$ (yield a goodness-of-fit $\chi^2 \sim 1.49$) as shown in Fig.~\ref{FIG1}. The 1:2 site ordering with propagation vector $\frac{1}{3}(111)$ in a pseudocubic structure forms a layered edged-shared triangular lattice of Cu$^{2+}$ in the (111) plane. The two successive triangular layers of Cu$^{2+}$ are separated by two layers of Ta$^{5+}$ as shown in the inset of Fig.~\ref{FIG1}. A similar structure is also found in Sr$_3$CuSb$_2$O$_9$~\cite{Kundu267202}. Note that the Cu$^{2+}$ moments form an edge-shared bilateral or quasi-equilateral triangle with a very small difference in the bond lengths (5.56, 5.70, and 5.70~\AA) along the edges of the triangle. The smallest bond length between inter-plane Cu$^{2+}$ ions is about $\sim 6.93$~\AA.


\begin{figure}
\includegraphics[scale=0.33]{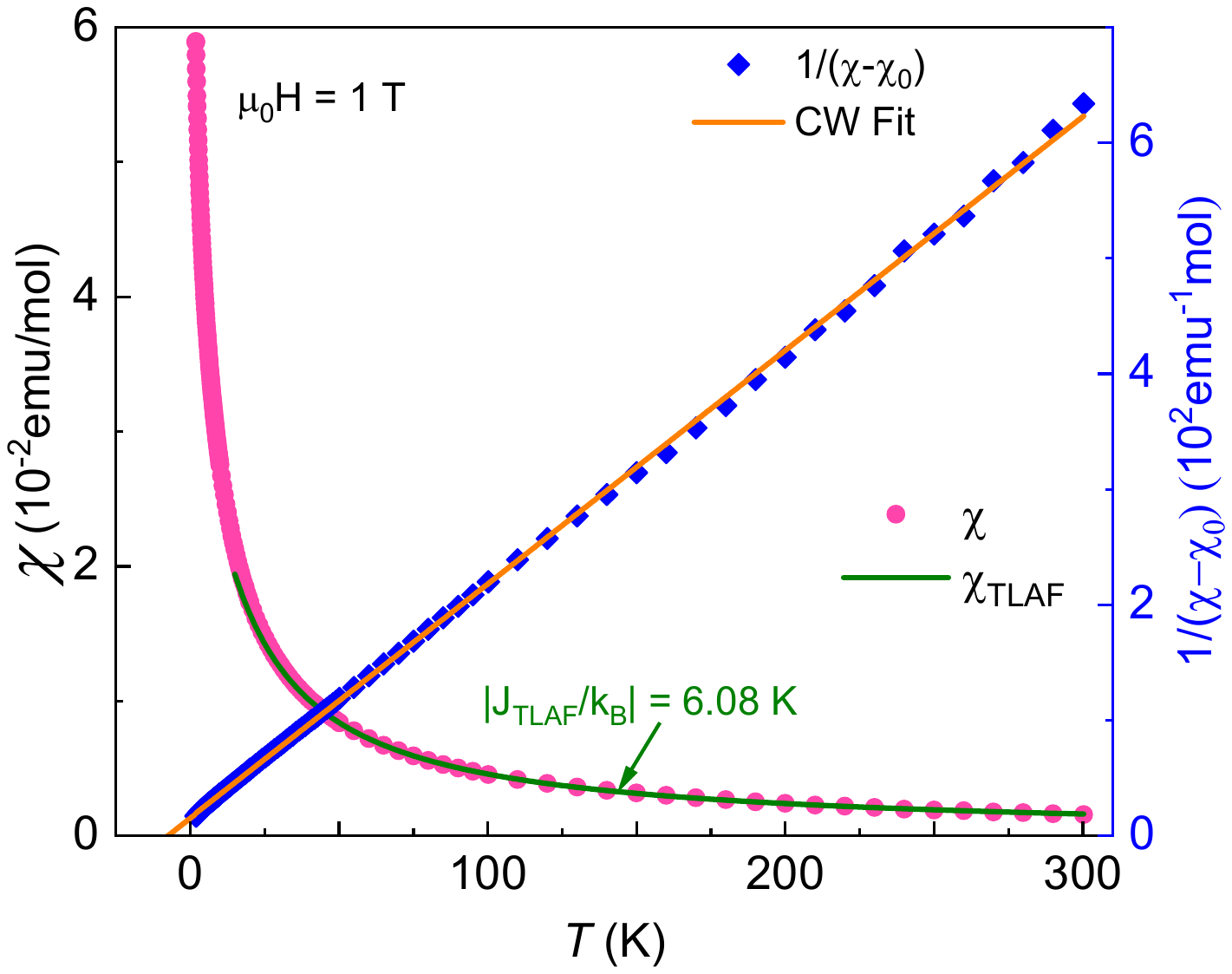}
\caption{\label{FIG2}Left $y$-axis: $\chi$ as a function of $T$ measured at $\mu_0H=1$~T. The solid line represents the fit using an anisotropic ($J_1-J_2$) $S = 1/2$ TLAF model [Eq.~\eqref{anisotropy_susceptibility}]. Right $y$-axis: inverse-susceptibility (after subtracting $\chi_0$) as a function of $T$ at $\mu_0H=1$~T and solid line is the Curie-Weiss fit.}  
\end{figure}

\textit{Magnetization:}
Temperature dependent dc magnetic susceptibility $\chi(T) (\equiv M/H$) measured at $\mu_0H=1$~T is depicted in Fig.~\ref{FIG2}. $\chi$ increases monotonically with lowering temperature and no anomaly associated with magnetic LRO is observed down to 1.85~K. $\chi(T)$ measured in zero-field-cooled and field-cooled protocols in a low magnetic field (50~Oe) shows no bifurcation down to 1.85~K~\cite{Supp}, ruling out a spin-glass like transition. In Fig.~\ref{FIG2}, we fitted the $\chi(T)$ data in high temperatures ($T > 100$~K) by the modified Curie-Weiss (CW) law [$\chi(T)=\chi_0 + C/(T-\theta_{\rm CW})$] that yields temperature independent susceptibility $\chi_0 \simeq - 2.18 \times 10 ^{-5}$~emu/mol-Cu$^{2+}$, Curie constant $C \simeq 0.49$~emu.K/mol-Cu$^{2+}$, and characteristic CW temperature $\theta_{\rm CW} \simeq -7.56$ $ \pm 0.11$~K. The negative sign of $\theta_{\rm CW}$ indicates dominant antiferromagnetic (AFM) interaction between Cu$^{2+}$ spins. From the value of $C$, the effective moment is estimated to be $\mu_{\rm eff}= \sqrt{3 k_{\rm B} C / N_{\rm A}} \simeq 1.97(8) \mu_{\rm B}$/Cu$^{2+}$ (where, $k_{\rm B}$, $N_{\rm A}$, $\mu_{\rm B}$, and $g$ are the Boltzmann constant, Avogadro’s number, Bohr magneton, and Land\'{e} $g$-factor, respectively). This value is indeed comparable with the expected value of $\mu_{\rm eff} \simeq 1.73~\mu_{\rm B}$/Cu$^{2+}$ [$= g\sqrt{S(S+1)}\mu_{\rm B}$] for Cu$^{2+}$ ($S=1/2$). Here, $\theta_{\rm CW}$ represents the overall energy scale of the exchange couplings and one can estimate the average intra-layer coupling ($J_{\rm CW}/k_{\rm B}$) as $\theta_{\rm CW} = - z J_{\rm CW} S (S+1)/3 k_{\rm B}$. Taking the experimental value of $\theta_{\rm CW}$ and the number of NN spins $z=6$ for a 2D-TLAF, we got $J_{\rm CW}/k_{\rm B} \simeq - 5.04$~K.

As inferred from the crystal structure, one expects a spatial anisotropy in the triangular unit due to a slightly varying bond length. Therefore, to analyze $\chi(T)$ we fitted the data in the high-temperature region ($T > 100$~K) by $\chi (T) = \chi_0 + \chi_{\rm TLAF}(T)$. $\chi_{\rm TLAF}$ is the expression of high-temperature series expansion (HTSE) for a $S=1/2$ 2D spatially anisotropic TLAF which has the form~\cite{Zheng134422}
\begin{equation}
\label{anisotropy_susceptibility}
\chi_{\rm TLAF} = \frac{N_{\rm A}g^2\mu_{\rm B}^2}{k_{\rm B}T} \times \sum_{\rm n=0} \left( \frac{J_2}{T} \right)^{\rm n} \sum_{\rm m=0}^{\rm n} \frac{c_{\rm m,n}y^{\rm m}}{4^{\rm n+1}{\rm n}!}.
\end{equation}
Here, $y=J_1/J_2$ with $J_1$ and $J_2$ representing the NN and NNN exchange interactions, respectively. The integer coefficients $c_{m,n}$ are given in Ref.~\cite{Zheng134422}. By fixing the $\chi_0$ value (obtained from CW fit), the fit yields $J_2/k_{\rm B} \simeq 6.08$ $\pm 0.35$~K, and $J_2/J_1 \simeq 1$. The value $J_2/J_1 \simeq 1$ indicates an effective equilateral triangular lattice as far as the strength of exchange interactions are concerned. Moreover, the obtained value of $J_2/k_{\rm B}$ is in good agreement with $J_{\rm CW}/k_{\rm B}$, further endorsing a nearly isotropic triangular lattice. We also fitted the $\chi(T)$ data by the HTSE of a $S=1/2$ isotropic TLAF that uses Pad\'{e} approximation~\cite{Supp}. The fit results in nearly identical exchange coupling, $J_{\rm iso}/k_{\rm B} \simeq 6.09$ \textcolor{blue}{$\pm 0.40$}~K.


\begin{figure}
\includegraphics[scale=0.34]{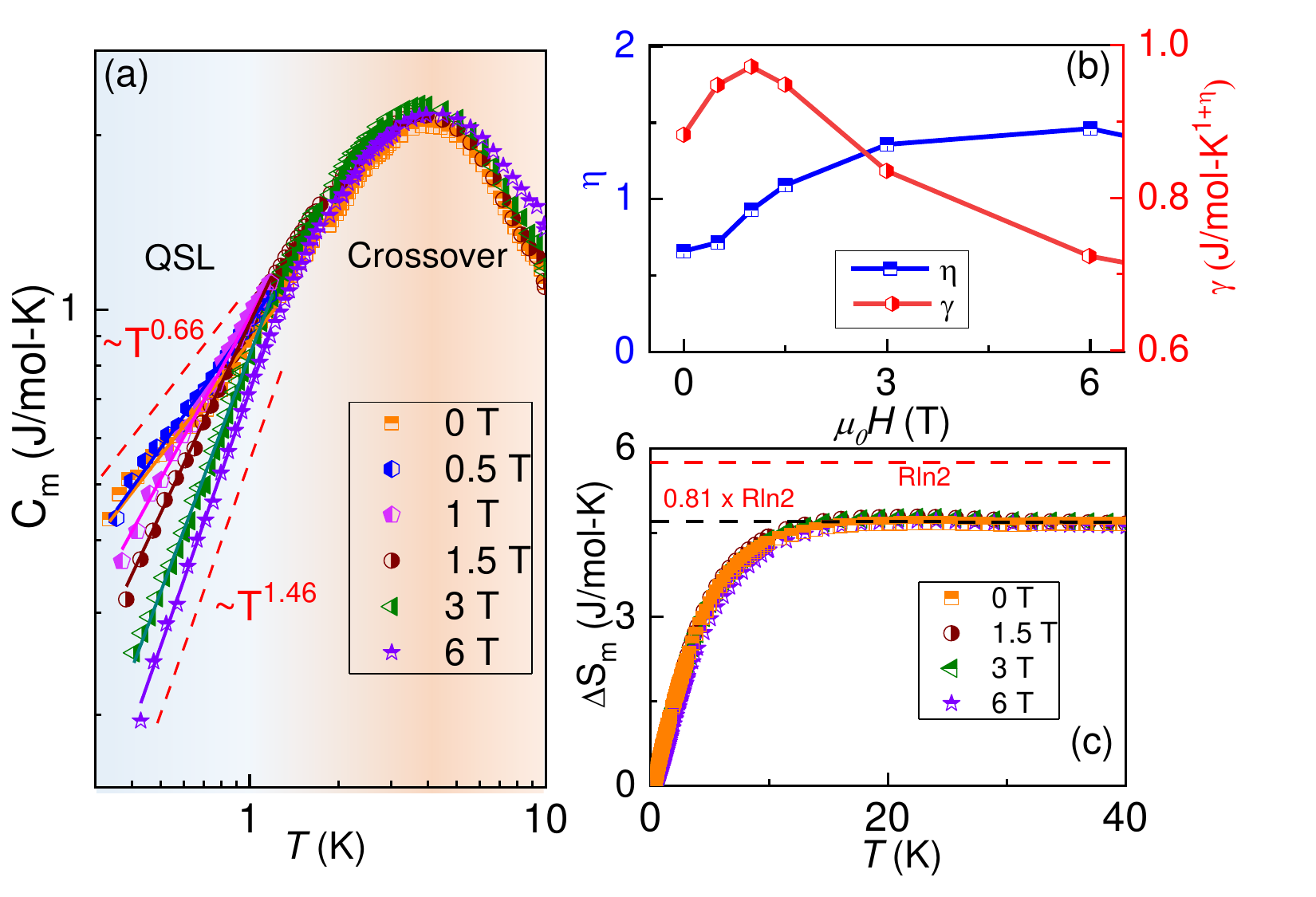}
\caption{\label{FIG3}(a) Magnetic specific heat ($C_{\rm m}$) as a function of $T$ for different fields. The solid lines are the fits using $C_{\rm m} \sim \gamma T^\eta$ and red dashed lines correspond to $T^{0.66}$ and $T^{1.46}$. (b) Variation of $\eta$ and $\gamma$ with the magnetic field. (c) Magnetic entropy ($\Delta S_{\rm m}$) as a function of $T$ and the dotted lines mark the theoretically expected and experimentally observed values of entropy.}  
\end{figure}

\textit{Specific Heat:} 
Specific heat not only provides information about magnetic LRO but also the low-energy excitations of a spin system. Specific heat data in zero-field as well as in applied fields show the absence of a $\lambda$-type anomaly reminiscent of a magnetic LRO down to 0.33~K. To extract the magnetic contribution to the specific heat ($C_{\rm m}$), we have subtracted the lattice contribution ($C_{\rm lat}$) from the total specific heat ($C_{\rm p}$). The lattice contribution was estimated considering both the Debye and Einstein models in the temperature range of $30 - 190$~K with one Debye and two Einstein terms (see SM)~\cite{Supp}. $C_{\rm m}$ vs $T$ presented in Fig.~\ref{FIG3}(a) features a broad hump at around $\sim 4$~K ($T_{hump}$), which indicates crossover from a thermally disordered paramagnet to a quantum paramagnetic QSL state, typically observed in QSL systems~\cite{Dey174411,Li16419}. Moreover, for a Heisenberg TLAF, such a hump is expected at $T_{hump}/J\simeq 0.55$~\cite{Elstner93}, which yields $J/k_B\approx 7.27$~K consistent with the $J$ evaluated from magnetization measurements. For $T\leq 1.2$~K, $C_{\rm m}$ measured in zero-field follows a power-law behavior $C_{\rm m} \sim \gamma T^\eta$ with $\eta = 2/3$. While for different gapless QSL candidates, $C_m$ follows either linear~\cite{PhysRevLett.106.147204, Mustonen2018} or quadratic behavior with the temperature (for Dirac spin liquid~\cite{Kundu267202, PhysRevB.105.L121109}), in contrast to that for SCTO $\eta = 2/3$ is quite rare, and interestingly, is indeed consistent with a theoretical prediction for an equilateral TLAF with the presence of gapless Fermi surface as $C_{\rm m} \sim k_{\rm B} \nu_0 t_{\rm spinon}^{1/3} (k_{\rm B}T)^{2/3}$. Here, $\nu_0$ is the density of states at the spinon Fermi surface, and $t_{\rm spinon}$ is the spinon hopping amplitude~\cite{Motrunich045105}. The obtained field dependence of $\eta$ and $\gamma$ are shown in Fig.~\ref{FIG3}(b). With the application of a magnetic field, the exponent $\eta$ grows slowly, probably due to the gradual suppression of 2D quantum correlations. We also calculated the magnetic entropy change ($\Delta S_{\rm m}$) for different fields as shown in Fig.~\ref{FIG3}(c). $\Delta S_{\rm m}$ recovers only $\sim 81$\% of R$ln2$ above 10~K. A signature of frustrated magnet is that the entropy releases over a broad range of temperatures, thus the remaining 19\% of entropy will be released at further low temperatures, which means persistence of strong correlation among the Cu$^{2+}$ spins are present at further low temperatures. Such a reduction in entropy has been observed in other frustrated magnets~\cite{Jana169814, Ranjith180401, Singh075128}.

In addition, we have also measured temperature-dependent thermal conductivity $\kappa(T)$ down to $T = 2$~K in different magnetic fields (see SM)~\cite{Supp}. The low-temperature data are fitted by $\kappa/T = a + bT^2$ that yields a non-zero intercept $a \simeq 0.0211$~mWK$^{-2}$cm$^{-1}$ which further corroborate a gapless QSL state with a finite spinon density of states. However, a detailed quantitative analysis and solid evidence of QSL from thermal conductivity measurements require data at ultra-low temperatures.
\begin{figure*}
\includegraphics[scale=0.48]{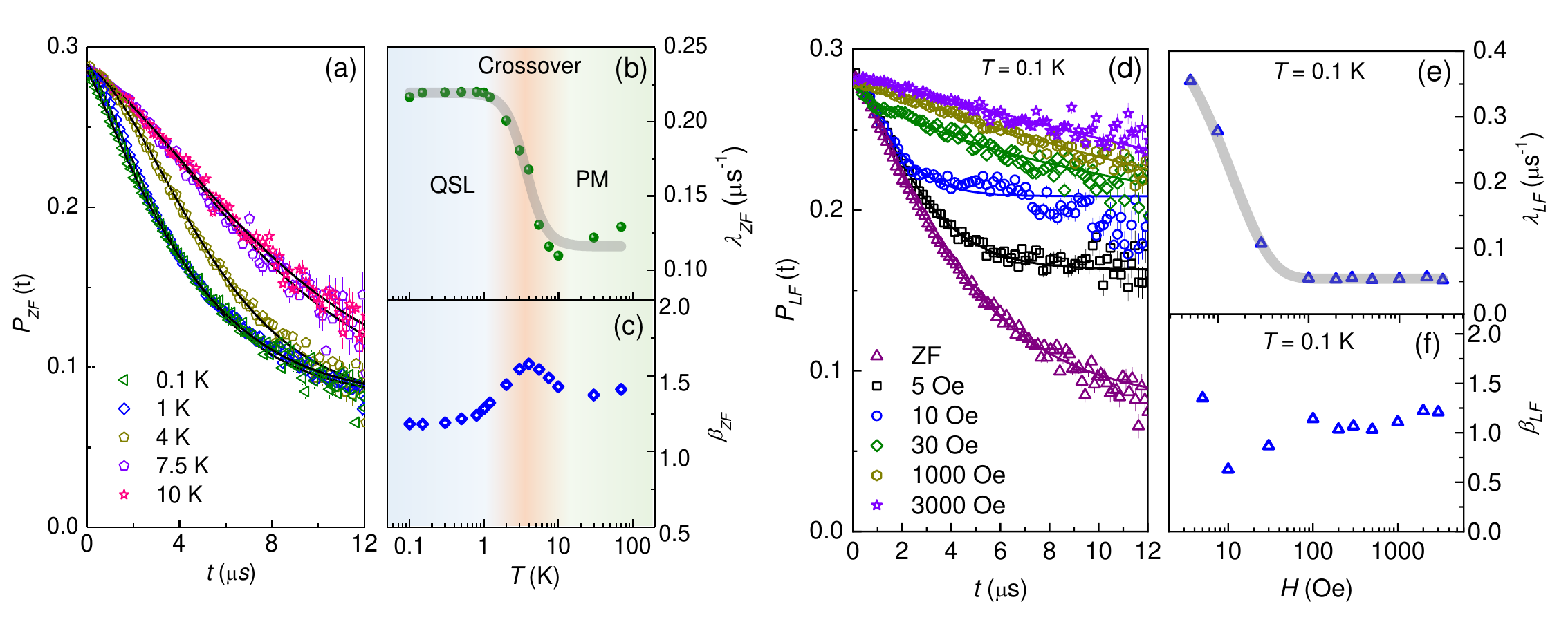}
\caption{\label{FIG4}(a) ZF-$\mu$SR asymmetry spectra as a function of $t$ for different temperatures. Solid lines are the fits using Eq.~\eqref{musr_eqn}. (b) ZF relaxation rate ($\lambda_{\rm ZF}$) vs $T$. (c) $\beta_{\rm ZF}$ as a function of $T$. The orange-shaded region represents the crossover region ($1.2 \leq T \leq 10$~K) from paramagnetic to the QSL state. (d) $\mu$SR asymmetry spectra measured in different longitudinal fields at $T = 0.1$~K and the solid lines represent the fits using Eq.~\eqref{musr_eqn}. (e) The corresponding relaxation rate $\lambda_{\rm LF}$ as a function of the longitudinal field. (f) $\beta_{\rm LF}$ as a function of the longitudinal field.}
\end{figure*}

\textit{Muon Spin Relaxation ($\mu$SR):} Being sensitive to a local internal magnetic field as small as 0.1~Oe, $\mu$SR is an ideal microscopic tool to probe magnetic LRO. In addition, $\mu$SR can distinguish between static and dynamic correlations among the spins making it a powerful tool to uncover the putative QSL behaviour. Hence, to elucidate the magnetic ground state of SCTO, we collected $\mu$SR data in zero-field (ZF) as a function of temperature as well as in longitudinal fields (LFs) at the base temperature of 0.1~K. In the following, we delineate our observations from the $\mu$SR data.

(i) ZF $\mu$SR asymmetries displayed in Fig.~\ref{FIG4}(a) confirm the absence of magnetic LRO down to 0.1~K, as they decay continuously without any oscillations or initial aymmetry drop [Fig.~\ref{FIG4}(a)]. The ZF $\mu$SR asymmetries are well-fitted by a stretch exponential function with an extra background term $B_{\rm bg}$ (owing to some muons missing the sample and placed at the Ag-sample holder and the cryostat wall.)~\cite{Li097201} as
\begin{equation}
\label{musr_eqn}
P(t) = P(0) \exp{(- \lambda t)^{\beta}} + B_{\rm bg}.
\end{equation}
Here, $ P(0) \sim 0.28 $ (weakly temperature and field dependent) is the initial asymmetry, $ \lambda$ is the relaxation rate, and $\beta$ is the stretching exponent. While fitting the ZF $\mu$SR asymmetries, $B_{\rm bg} \approx 0.08$ is kept constant for all the temperatures. The obtained fitting parameters ($\lambda_{\rm ZF}$ and $\beta_{\rm ZF}$) are plotted in Fig.~\ref{FIG4}(b and c).

(ii) In the high temperature ($T > 10$~K) paramagnetic regime, the uncorrelated Cu$^{2+}$ spins fluctuate rapidly and randomly. The fluctuation rate can be calculated using the strength of exchange coupling ($J/k_{\rm B} \simeq 5.04$~K) as $\nu = \sqrt{z}Js/\hbar \sim 1.6 \times 10^{10}$~Hz \cite{Lee024413,Uemura3306}, where $\nu$ is the spin fluctuation rate and $z = 6$ is the coordination number for a 2D TLAF. $\lambda_{\rm ZF}$ (at $T > 10$~K) can provide an idea about the local internal field distribution ($\Delta$) based on the Redfield formula~\cite{MllerWarmuch1991}, $\lambda(T\geq 10~K, H) = 2 \Delta^2 \nu /(\nu^2 + \gamma^2_\mu \mu^2_0 H^2)$, which yields $\Delta \sim 5.1 \times 10^7$~Hz. This confirms the fast-fluctuation limit ($\Delta \ll \nu$)~\cite{Bono187201,Uemura3306} is responsible for the temperature-independent relaxation rate in the high-temperature range. With lowering the temperature, the relaxation rate $\lambda_{\rm ZF}$ starts increasing below $T \leq 10$~K, which implies slowing down of fluctuating moments due to the growth of correlations among the Cu$^{2+}$ spins~\cite{Bono187201,Keren107204}. This temperature regime is marked as the crossover region from paramagnetic to QSL state in Fig.~\ref{FIG4}(b), consistent with the broad hump observed in $C_{\rm m}(T)$ [see Fig.~\ref{FIG3}(a)]. With further lowering in temperature, $\lambda_{\rm ZF}$ becomes temperature independent (for $T \leq 1.2$~K), indicating the persistence of spin dynamics. This persistent spin dynamics is considered to be a generic feature of QSL, as reported for other celebrated QSL candidates~\cite{Li097201,Kundu267202,Dey174411,Zhang085115}. It is interesting to note that the saturation of $\lambda_{\rm ZF}$ below $\sim 1.2$~K exactly coincides with the temperature range where $C_{\rm m}(T)$ follows a power-law behavior. Thus, the behavior of $\lambda_{\rm ZF}(T)$ along with the power-law dependent of $C_{\rm m}(T)$ provides a robust signature of QSL at low temperatures.


(iii) LF $\mu$SR experiments are performed to explore the nature of the spin dynamics at low temperatures. LF-$\mu$SR asymmetry spectra measured in different fields at $T = 0.1$~K are shown in Fig.~\ref{FIG4}(d). They are fitted well by Eq.~\eqref{musr_eqn}. The obtained field dependence of $\lambda_{\rm LF}$ and $\beta_{\rm LF}$ are depicted in Fig.~\ref{FIG4}(e) and (f), respectively. The quick suppression of $\lambda_{\rm LF}$ below 100~Oe reflects the decoupling of the nuclear contribution to the relaxation rate~\cite{Kermarrec100401,Zhang085115}. More interestingly, above 0.01~T, the remaining $\lambda_{\rm LF} \sim 0.05~\mu s^{-1}$, originating from the electronic contribution, is almost field independent up to a LF of 3000~Oe. At low temperatures ($T \leq 1.2$~K), if we assume that the relaxation process in the plateau regime of $\lambda_{\rm ZF}$ is due to a static local field then it would correspond to $\sim 0.25$~mT (as $B_{\rm loc}=\lambda / \gamma_\mu$, where $\gamma_\mu = 2 \pi \times 135.5~s^{-1}\mu T^{-1}$ is the gyromagnetic ratio of muons). In such a scenario, an LF of 750~Oe, which is 5 times larger than $B_{\rm loc}$ would be sufficient to decouple the relaxation channel. On the contrary, even in an LF of 3000~Oe, the decoupling of the relaxation channel was not achieved. One may require higher LFs to completely decouple the relaxation channel~\cite{ishant2024}. This observation demonstrates that the correlation among the spins is entirely dynamic (not static) in nature, as expected for a QSL state~\cite{Kundu267202,Sarker241116,Li097201,Balz942,Tripathi064436}.




(iv) Typically, in a spin-glass state, the value of $\beta$ is predicted to be about 1/3~\cite{Ogielski7384,Campbell1291}. On the contrary, the obtained $\beta$ value shown in Fig.~\ref{FIG4}(c) attains a constant value of $\sim 1.2$ below the crossover region. Furthermore, the magnitude of $\lambda_{\rm ZF}$ ($\sim 0.12~\mu s^{-1}$ at $T \geq 10 $~K) becomes double ($\sim 0.22~\mu s^{-1}$ at $T \leq 1 $~K) upon cooling the system below the crossover regime, which is also in contrast to that expected for a spin-glass type transition, where the relaxation rate should increase by few orders of magnitude~\cite{Uemura546,Uemura3306}. These observations rule out the possibility of a spin-glass state.

It is important to highlight that the system doesn't order down to $T_{\rm min} \sim 0.1$~K that sets the lower limit of the frustration parameter $f = \theta_{\rm CW}/T_{\rm min} \simeq 76$, characterizing SCTO a highly frustrated magnet. Specific heat and $\mu$SR results establish a highly dynamic ground state with spinon excitations and a footprint of gapless QSL. The analysis of magnetic susceptibility suggests that the system can be treated as an isotropic TLAF with Heisenberg interactions. In such a scenario, the ground state is expected to be a 120$^{\circ}$ N$\acute{e}$el order rather than a QSL state, if only the NN couplings are considered. Further, despite partial site occupancy, the Cu and Ta sites are arranged periodically with 1:2 order, forming separate layers of Cu$^{2+}$ with a triangular lattice. This ensures the minimal effect of disorder driving QSL state. Note that, systems with disorder may stabilize random-singlet-state, which exhibits scaling behavior in their physical properties as observed in different compounds with site disorder~\cite{PhysRevResearch.2.013099, PhysRevB.106.174406, Kimchi2018}. However, as expected the physical properties of SCTO do not show any such scaling behavior. All these observations reflect the possible role of NNN interaction ($J_{\rm nn}$) or inter-layer interaction ($J_{\perp}$) [inset (b) of Fig.~\ref{FIG1}] in stabilizing the gapless QSL state. Nevertheless, the inelastic neutron scattering experiments on a good-quality single crystal would be essential to shed light on this issue.

\textit{Conclusion:} In summary, our studies divulge that the Cu$^{2+}$ and Ta$^{5+}$ ions feature a precisely calibrated 1:2 site ordering, resulting in seperate planes of Cu$^{2+}$ moments with propagation vector $k = \frac{1}{3}(111)$ which creates effective equilateral triangular lattice. Magnetic susceptibility data agree well with the isotropic $S=1/2$ TLAF model with a leading exchange coupling of $J/k_{\rm B} \simeq 6.09$~K.
Specific heat data suggest the absence of magnetic LRO down to 0.33~K and provide evidence for spinon excitations. The absence of magnetic LRO was further corroborated by the $\mu$SR data measured down to 0.1~K, setting a very high degree of magnetic frustration ($f > 76$) in SCTO. From the $\mu$SR analysis, the ground state is found to be a highly dynamic state with no static order, a hallmark of QSL. Thus, our detailed investigation unambiguously established that SCTO is one of the rare examples of a gapless QSL realized in a $S = 1/2$ Cu$^{2+}$-based Heisenberg TLAF. The onset of such a gapless QSL is anticipated to be due to the complex interplay of different exchange couplings (NN and interactions beyond NN). We believe that our results would instigate further experimental as well as theoretical investigations to settle the origin of the emergence of QSL in SCTO.

\textit{Acknowledgment:} K.B. and M.M. would like to thank the Department of Science and Technology, India for the access to the experimental facility and financial support for the experiment conducted at ISIS muon source \cite{SCTO} and Jawaharlal Nehru Centre for Advanced Scientific Research (JNCASR) for managing the project. 
S.M. and R.N. would like to acknowledge SERB, India, bearing sanction Grant No.~CRG/2022/000997. K.B. and M.M. also thank I. Ishant for fruitful discussions.

\bibliography{references}
\end{document}